\documentclass{webofc}
\usepackage[varg]{txfonts}   
\begin{document}
%
\title{Azimuthal anisotropy measurement of multi-strange \\hadrons in Au+Au collision at $\sqrt{s_{NN}}$ = 27 and 54.4 GeV \\at STAR }
%
%

\author{\firstname{Prabhupada} \lastname{Dixit (for the STAR collaboration)}\inst{1}\fnsep\thanks{\email{dixitprabhupada471@gmail.com}} 
}

\institute{Indian Institute of Science Education and Research, Berhampur, 760010, India
          }
      
\abstract{%
  In these proceedings, we present the second order ($v_{2}$) and the third order ($v_{3}$) azimuthal anisotropy measurements of strange and multi-strange hadrons such as $K^{0}_{S}$, $\Lambda$, $\bar{\Lambda}$, $\phi$, $\Xi^{-}$, $\bar{\Xi}^{+}$, $\Omega^{-}$, $\bar{\Omega}^{+}$ in Au+Au collisions at $\sqrt{s_{NN}}$ = 27 and 54.4 GeV at STAR. The centrality dependence of $v_{2}$ and $v_{3}$ as well as the number of constituent quark scaling (NCQ scaling) for all the particles mentioned above are discussed. The ratio of $v_{2}$ of $\phi$ mesons to that of $\bar{p}$ is presented at $\sqrt{s_{NN}}$ = 54.4 GeV which gives a hint of mass ordering violation at low $p_{T}$.
}
%
\maketitle
%
\section{Introduction}
\label{intro}
Relativistic heavy-ion collision is a unique tool to study matter at the extreme conditions of temperature and density. The collective expansion of the medium formed in heavy-ion collision is called flow. Depending on the initial spatial configuration of the colliding system and fluctuations, there can be the presence of different orders of anisotropy in the azimuthal distributions, known as anisotropic flow. In non-central nucleus-nucleus collisions, the overlap region of two colliding nuclei looks like the shape of an almond and in this region the pressure gradient is different along different axis of the system. The difference in pressure gradient leads to momentum anisotropy which is the dominant cause for the development of elliptic flow.
But the event-by-event fluctuations of the nucleons in the colliding nuclei can give rise to other types of flow such as triangular flow, rectangular flow, etc. All these flow coefficients can be studied by using Fourier series expansion of the azimuthal distribution of particles given by
\begin{equation}
\label{eq-1}
E\frac{d^{3}N}{dp^{3}} =\frac{1}{2\pi}\frac{d^{2}N}{p_{T}dp_{T}dy}\left[1 + \sum_{n} 2v_{n}\cos n(\phi-\Psi_{R})\right].
\end{equation}
Here $v_{2}$ and $v_{3}$ represent elliptic flow coefficient and triangular flow coefficient respectively. The $n^{th}$ order flow coefficient is given by
\begin{equation}
\label{eq-2}
v_{n} = \langle \cos n(\phi -\psi_{n})\rangle.
\end{equation}
Here $\phi$ is the azimuthal angle of the particles and $\psi_{n}$ is the $n^{th}$ order event plane angle.
The simultaneous measurement of $v_{2}$ and $v_{3}$ is important to constrain the shear viscosity over entropy density ratio ($\eta/s$)~\cite{Ref1}. In this study, we have used strange and multi-strange hadrons because of their small hadronic interaction cross-section and early freeze-out~\cite{Ref2}, the measured flow coefficients remain least affected in the hadronic phase.


\section{Analysis method}
\label{sec-2}
The $n^{th}$ order event plane is given by
\begin{equation}
\psi_{n} =\frac{1}{n} \tan^{-1}\Big(\frac{\sum_{i} w_{i} \sin(n\phi_{i})}{\sum_{i} w_{i} \cos(n\phi_{i})}\Big),
\end{equation}
where $\phi_{i}$ is the azimuthal angle of the particle track and $w_{i}$ is the weight factor equal to $p_{T}$ $\times$ $\phi$-weight. The factor $\phi$-weight is the correction factor for azimuthal acceptance of the detector. Due to the finite resolution of the  event plane, the $v_{n}$ needs to be corrected by event plane resolution given by
\begin{equation}
R_{n}= \sqrt{\langle \cos n(\psi_{A}- \psi_{B})\rangle} .
 \end{equation}
Here $\psi_{A}$ and $\psi_{B}$ are two sub-event planes in -1$ < \eta < $-0.05 and 0.05 $< \eta < $1 regions respectively. In Eq.~\ref{eq-2}, the particle's $\phi$ and the event plane $\psi_{n}$ are taken from opposite $\eta$ region.


Invariant mass method~\cite{Ref3} is used to calculate $v_{n}$. A detailed procedure to apply this method can be found in Ref.~\cite{Ref8}.

\section{Results and discussion}
\label{sec-3}
\begin{figure}[h]
\centering
\includegraphics[width=9cm,clip]{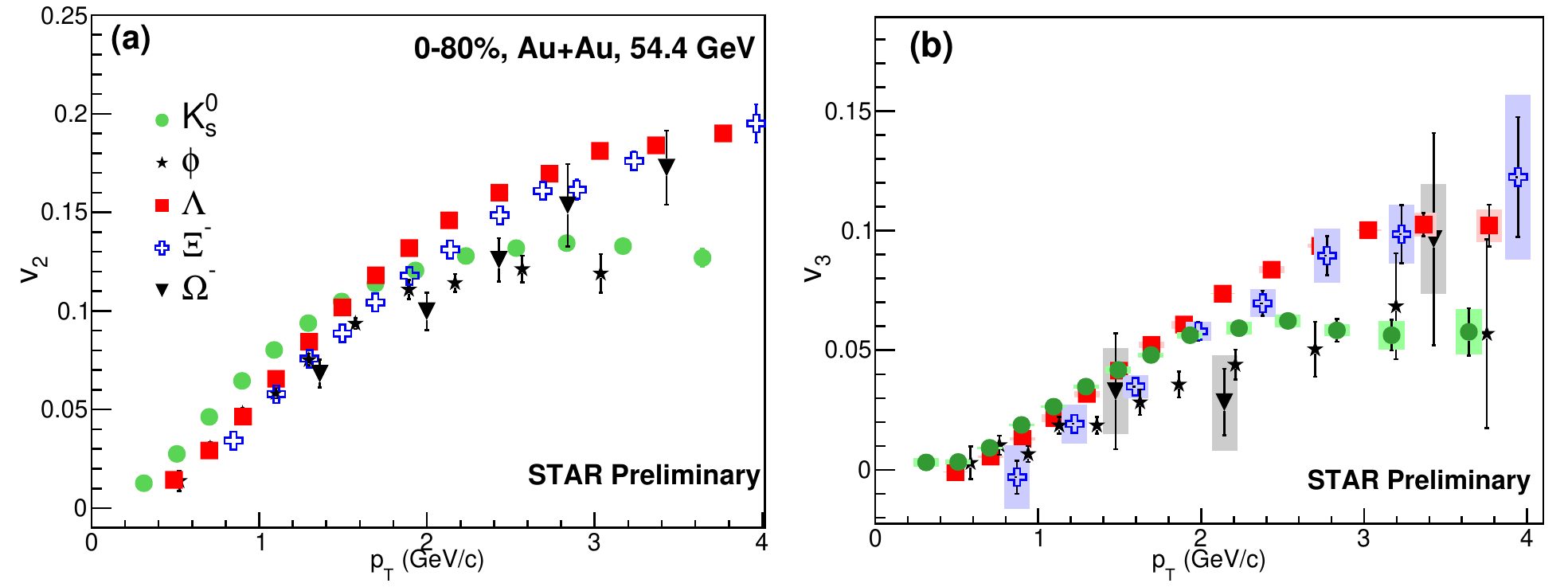}
\caption{ Panel-a shows $v_{2}$ and panel-b shows $v_{3}$ as a function of $p_{T}$ for $K^{0}_{S}$, $\Lambda$, $\phi$, $\Xi^{-}$, $\Omega^{-}$ in 0-80\% centrality. The vertical lines represent the statistical error bars and the shaded boxes represent the systematic error bars.}
\label{fig-3}       
\end{figure}

\begin{figure}[h]
\centering
\includegraphics[width=8cm,clip]{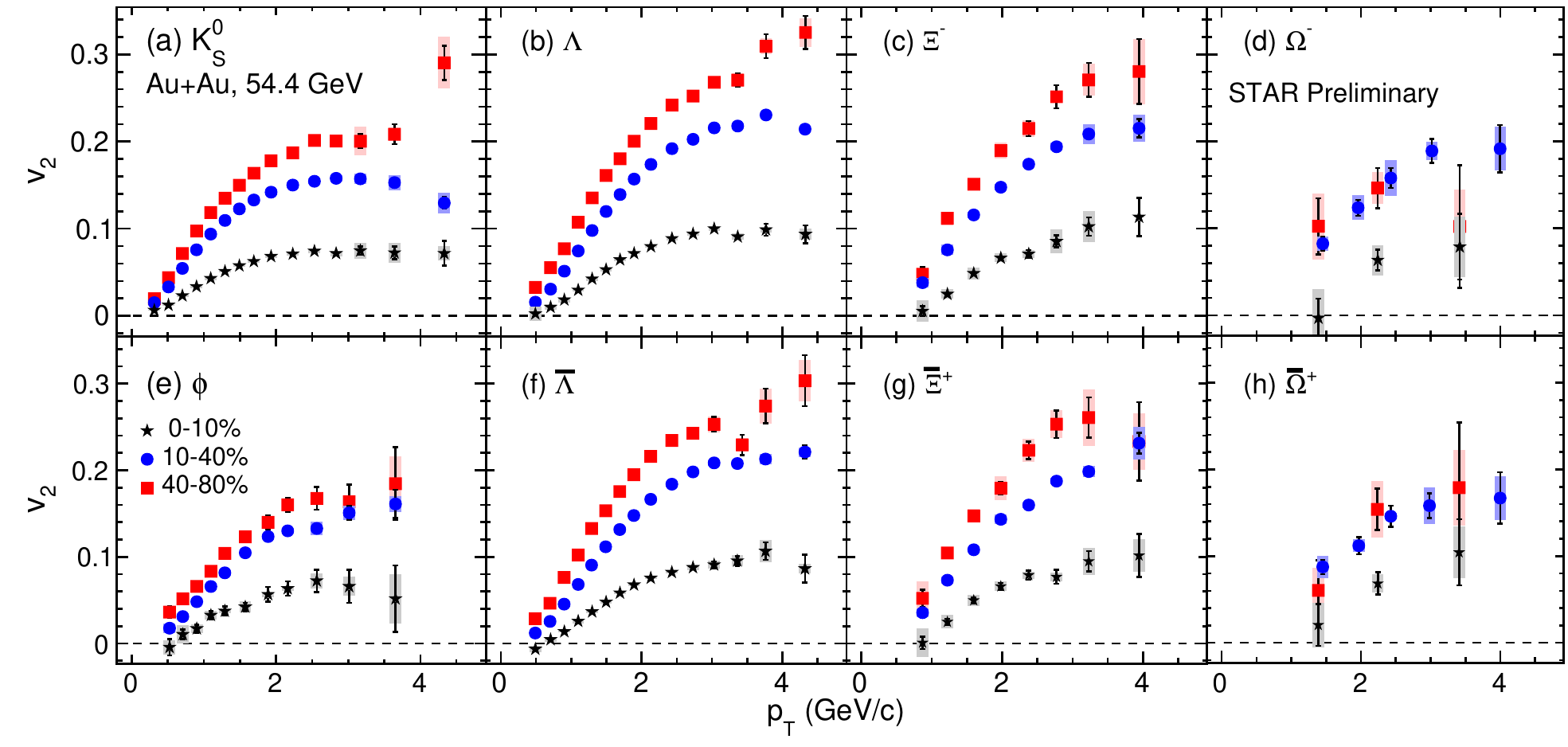}
\caption{$v_{2}$ is plotted as a function of $p_{T}$ for 0-10\%, 10-40\% and 40-80\% centrality classes. The vertical lines represent the statistical error bars and the shaded boxes represent the systematic error bars.}
\label{fig-4}       
\end{figure}

\begin{figure}[h]
\centering
\includegraphics[width=8cm,clip]{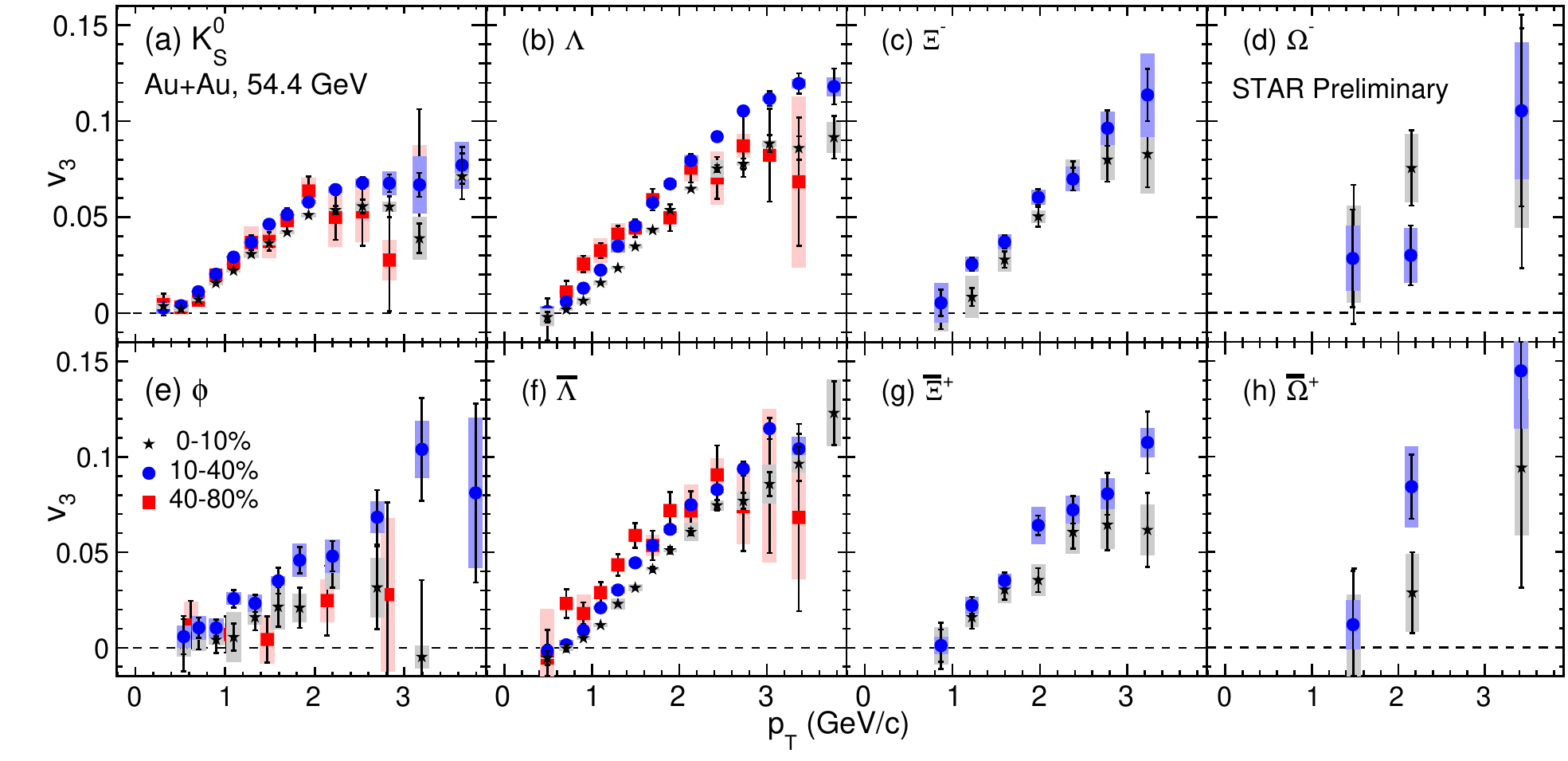}
\caption{ $v_{3}$ is plotted as a function of $p_{T}$ for 0-10\%, 10-40\% and 40-80\% centrality classes. The vertical lines represent the statistical error bars and the shaded boxes represent the systematic error bars. }
\label{fig-5}       
\end{figure}

Figure~\ref{fig-3} shows $v_{2}$ and $v_{3}$ plotted as a function of $p_{T}$ for $K^{0}_{S}$, $\Lambda$, $\phi$, $\Xi^{-}$, $\Omega^{-}$  at $\sqrt{s_{NN}}$ = 54.4 GeV in 0-80\% centrality. In low $p_{T}$ both $v_{2}$ and $v_{3}$ show a mass ordering which is due to the radial flow of the system~\cite{Ref4}, but at $p_{T}$ above 2 GeV/c a separation between baryons and mesons is observed which can be explained by the quark recombination model of hadronization~\cite{Ref5}. 
A centrality dependence study for both $v_{2}$ and $v_{3}$ is shown in Fig.~\ref{fig-4} and Fig.~\ref{fig-5} in three different centrality classes, 0-10\%, 10-40\% and 40-80\%. The elliptic flow $v_{2}$ shows a strong centrality dependence  because of initial spatial anisotropy dominance, whereas $v_{3}$ does not because it arises mostly due to event-by-event fluctuations in the initial state.

\begin{figure}[h]
\centering
\includegraphics[width=8cm,clip]{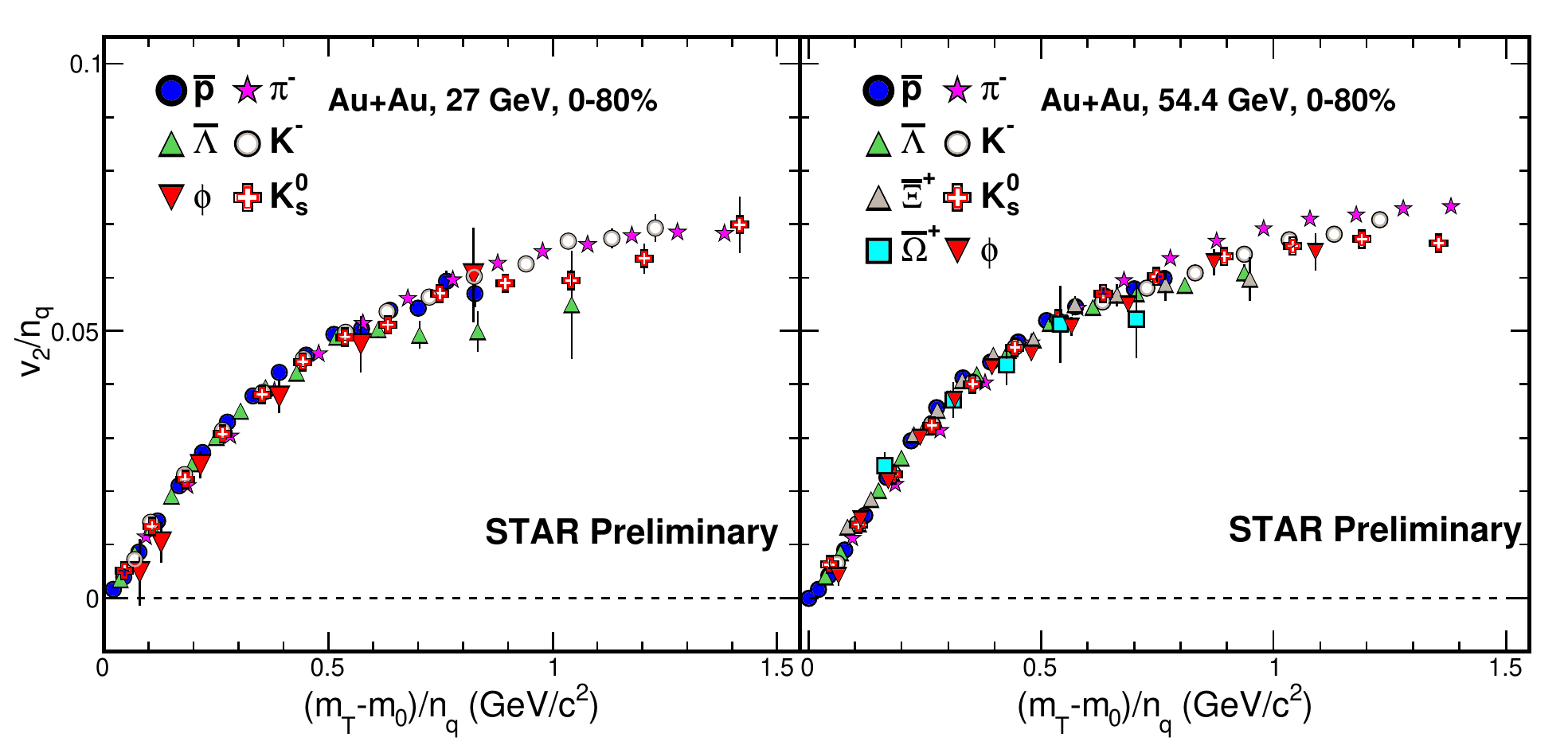}
\caption{NCQ scaled $v_{2}$ is plotted as a function of NCQ scaled transverse kinetic energy for 0-80\% centrality for $\sqrt{s_{NN}}$ = 27 (left) and 54.4 GeV (right). }
\label{fig-6}       
\end{figure}

\begin{figure}[h]
\centering
\includegraphics[width=6cm,clip]{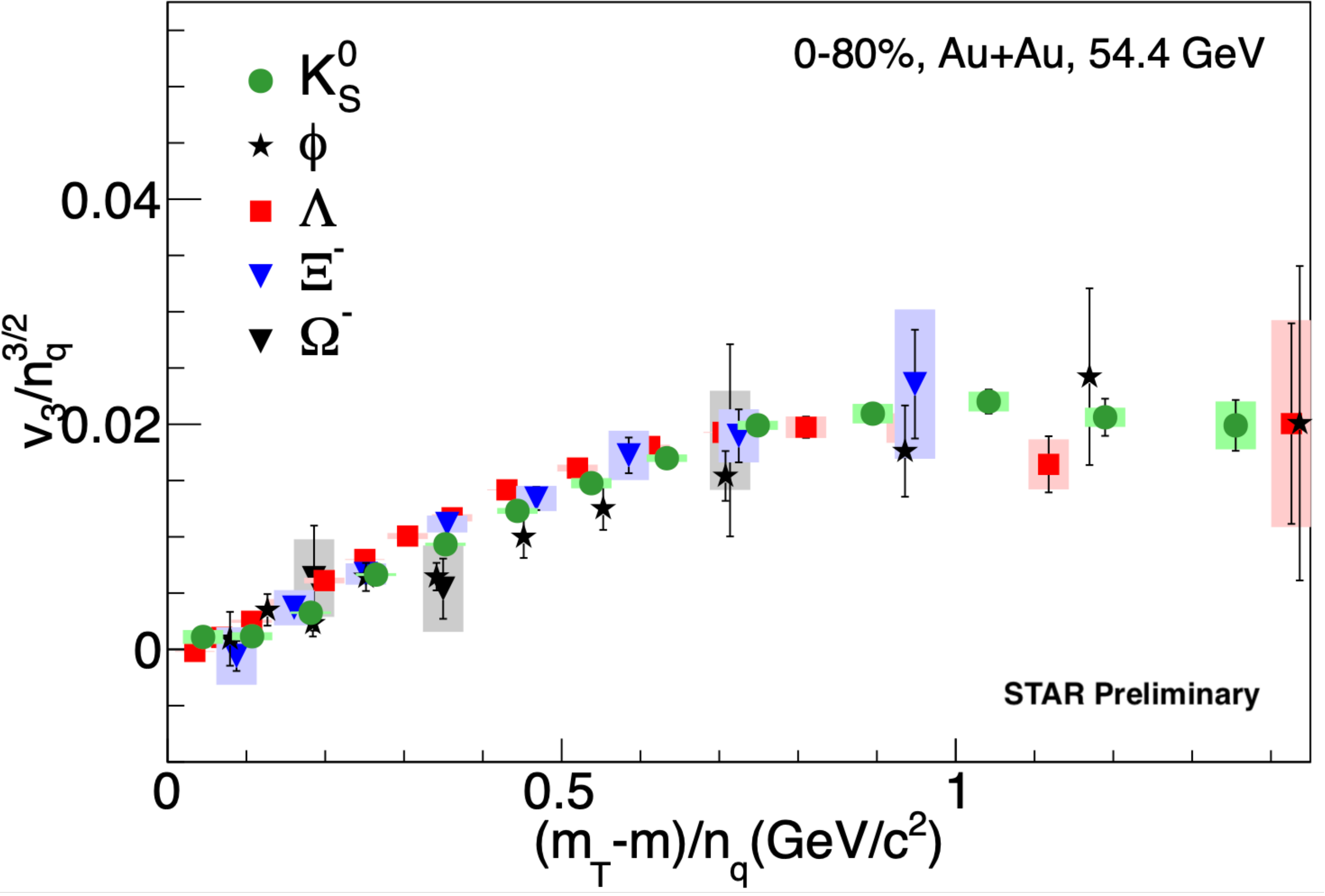}
\caption{NCQ scaled $v_{3}$ is plotted as a function of NCQ scaled transverse kinetic energy for 0-80\% centrality for  $\sqrt{s_{NN}}$ = 54.4 GeV. The vertical lines represent the statistical error bars and the shaded boxes represent the systematic error bars. }
\label{fig-10}       
\end{figure}

The NCQ scaled $v_{2}$ is studied as a function of NCQ scaled transverse kinetic energy for 0-80\% centrality at $\sqrt{s_{NN}}$ = 27 and 54.4 GeV. The NCQ scaled $v_{2}$ for all the particles shown in Fig.~\ref{fig-6} is found to fall on a single curve which suggests that the collectivity is developed during the QGP stage of the collision. 
The same result is found for $v_{3}$ also as shown in Fig.~\ref{fig-10}.

\begin{figure}[h]
\centering
\includegraphics[width=8cm,clip]{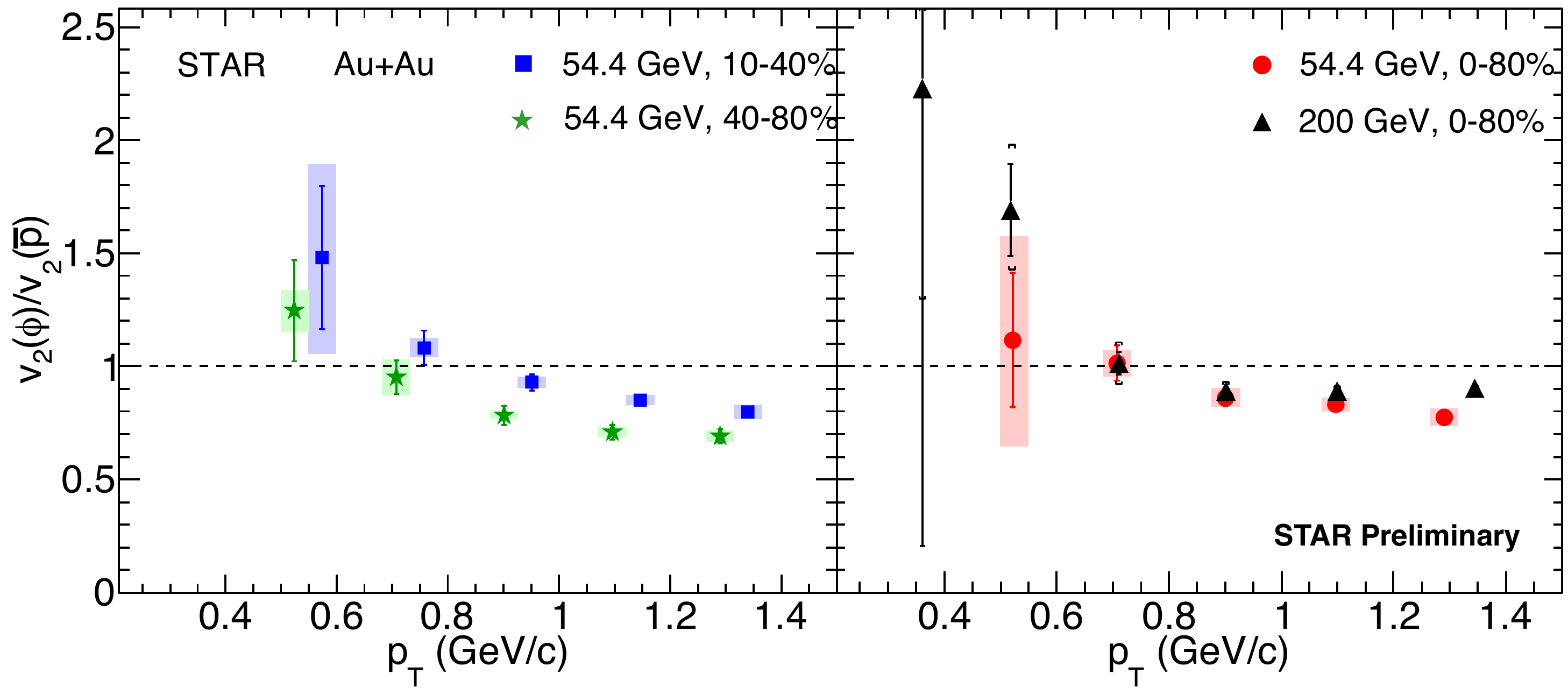}
\caption{The left panel shows the ratio of $\phi$ meson $v_{2}$  to $\bar{p}$ $v_{2}$ in Au+Au collision at $\sqrt{s_{NN}}$ = 54.4 GeV in 10-40\% and 40-80\% centralities. The right panel shows the comparison of the ratio with 200 GeV data points for minimum bias events. The vertical lines represent the statistical error bars and the shaded boxes represent the systematic error bars.}
\label{fig-7}       
\end{figure}

The ratio of $\phi$ meson $v_{2}$  to $\bar{p}$ $v_{2}$ is calculated as a function of $p_{T}$ as shown in Fig.~\ref{fig-7}. Since the mass of the $\phi$ meson (1.019 GeV/$c^{2}$) is greater than that of $\bar{p}$ (0.938 GeV/$c^{2}$), according to mass ordering at low $p_{T}$ the ratio should have been less than one. But we observe a hint of mass ordering violation at low $p_{T}$ which can be explained by larger hadronic rescattering effect on $\bar{p}$ compared to $\phi$ meson~\cite{Ref6}. A similar result is found in previous STAR measurement in minimum biased Au+Au collisions at $\sqrt{s_{NN}}$ = 200 GeV~\cite{Ref7}.

\section{Summary}
\label{sec-4}
Measurements of $v_{2}$ and $v_{3}$ as a function of $p_{T}$ of strange and multi-strange hadrons, $K^{0}_{S}$, $\Lambda$, $\bar{\Lambda}$, $\phi$, $\Xi^{-}$, $\bar{\Xi}^{+}$, $\Omega^{-}$, $\bar{\Omega}^{+}$ is presented in Au+Au collisions at midrapidity for 0-80\% centrality. A mass ordering is observed in the low $p_{T}$. At high $p_{T}$ ($ > $ 2 GeV/c), we observe a baryon to meson separation. A strong centrality dependence is observed for $v_{2}$, but the centrality dependence is absent for $v_{3}$. The NCQ scaling is found to hold for all the particles at $\sqrt{s_{NN}}$ = 27 and 54.4 GeV. The ratio of $v_{2}$ of $\phi$ meson to that of $\bar{p}$ gives a hint of mass ordering violation at low $p_{T}$.



%
%
%

\end{document}